\documentclass[a4paper,10pt]{article}
\usepackage{amsmath,graphicx}

\newcommand{\pni}{\par\noindent}

\begin{document}
\begin{center}
\textbf{BRANS-DICKE WORMHOLE AND SPACE-TIME FOAM}\\\vspace{0.3in}
\small{A. G. AGNESE AND M. LA CAMERA\\ \vspace{0.02in}
\emph {Dipartimento di Fisica dell'Universit\`a di 
Genova\\Istituto Nazionale di Fisica Nucleare, Sezione di 
Genova\\Via Dodecaneso 33, 16146 Genova, Italy}}\\ 
\emph{E-mail: agnese@ge.infn.it ;  lacamera@ge.infn.it} 
\end{center}
\begin{center}
\begin{tabular}{p{3in}}
\footnotesize{We introduce, by means of the Brans-Dicke scalar 
field, space-time fluctuations at scale comparable to Planck 
length near the event horizon of a black hole and examine their 
dramatic effects.}
\end{tabular}
\end{center} 
\vspace{0.2in}
\baselineskip = 1.1\baselineskip
\pni
It is generally believed that the space-time continuum will be 
replaced by a more fundamental level of description at length 
scales smaller than the Planck length $l_P$, where the 
classical description breaks down to a ``quantum foam'' [1] and 
the space-time topology fluctuates over a spatial distance 
comparable with~$l_P$. \par 
Moreover in recent papers 
Amelino-Camelia [2] and Ng and van Dam [3], basing on a gedanken 
timing experiment originally devised by Salecker and Wigner [4], 
argue that such a distance could be much greater that Planck 
length. As a consequence quantum gravity effects could be probed 
with current or future interferometers designed for the 
gravitational-wave detectors.\par In this communication, starting
from a different point of view, we give some arguments which can 
lead, already when distances vary at the Planck scale, to 
dramatic effects near the event horizon of the Schwarzschild 
vacuum solution now perturbed by the corresponding fluctuations 
of space-time. We will show that the metric breaks down, even for
large masses, in correspondence of the surface which should have 
been the event horizon, so it becomes unpossible not only to 
define an event horizon but even to foresee which singularity has
replaced it. These facts signal the need of a fully quantum 
description of gravity in the presence of distances suitable for 
the Planck regime even if surfaces are much greater than $l_P^2$ 
and the gravitational field in their neighbourhood is weak.\par 
To achieve this goal we start by considering the static 
spherically symmetric vacuum solution of the Brans-Dicke theory 
of gravitation [5]. \par The related calculations were performed 
by us in Ref. [6], working in the Jordan frame, where the action 
is given (in units $G_0 = c = 1$) by \begin{equation} S = 
\dfrac{1}{16 \pi }\,\int\, d^4x \sqrt{-\, g}\left[ \Phi R - \, 
\dfrac{\omega}{\Phi}\nabla^\alpha \Phi \nabla_\alpha \Phi \right]
\end{equation} and with a suitable choice of gauge. Here we quote
only the results relevant for the following.\par
The line element can be written as
\begin{equation}
ds^2 = e^{\mu(r)} dr^2 +  R^2(r) d\Omega^2 - e^{\nu(r)} dt^2
\end{equation}
where $d\Omega^2 = d\vartheta^2 + \sin^2 \vartheta d\varphi^2$ 
and, in the selected gauge:
\begin{align*}
R^2(r) = r^2 \left[1-\dfrac{2\eta}{r}\right]^{1-\gamma 
\sqrt{2/(1+\gamma)}} \tag{3a} \\ {} \\
e^{\mu(r)} = \left[ 1 - \dfrac{2\eta}{r}\right]^{-\gamma 
\sqrt{2/(1+\gamma)}}  \tag{3b} \\ {} \\
e^{\nu(r)} = \left[1 - 
\dfrac{2\eta}{r}\right]^{\sqrt{2/(1+\gamma)}} \tag{3c}
\end{align*}
\setcounter{equation}{3}
Here $\gamma$ is the post-Newtonian parameter 
\begin{equation}
\gamma = \dfrac{1+\omega}{2+\omega}
\end{equation}
and
\begin{equation}
\eta = M \sqrt{\frac{1+\gamma}{2}}
\end{equation}
Finally  the scalar field is given by
\begin{equation}
\Phi(r) = \Phi_0 \left[1-\dfrac{2\eta}{r}\right]^{(\gamma 
-1)/\sqrt{2(1+\gamma)}}
\end{equation}
while the effective gravitational coupling  $G(r)$  equals
\begin{equation}
G(r) = \dfrac{1}{\Phi(r)}\,\dfrac{2}{(1+\gamma)} 
\end{equation}
the factor $2/(1+\gamma)$ being absorbed, as in Ref. [5], in the 
definition of $G$.\par
Departures from Einstein's theory of General Relativity appear 
only if $\gamma \neq 1$, a possibility consistent with 
experimental observations  which estimate it in the range \, 
$1-0.0003 < \gamma < 1+0.0003$ corresponding to the dimensionless
Dicke coupling constant $|\omega|> 3000$.\par When $\gamma <1$ 
and $r \to 2\eta$, then $R(r)$, $e^{\nu(r)}$ and $G(r)$ go all 
to zero. Therefore we have a singularity with infinite red-shift 
and gravitational interaction decreasing while approaching the 
singularity.\par When $\gamma > 1$, the null energy condition 
(NEC) is violated [7] and a wormhole solution is obtained [6] 
with throat at 
\begin{equation}
r_0 = \eta \left[1+\gamma\,\sqrt{\dfrac{2}{1+\gamma}}\,\right]
= M \left[ \gamma + \sqrt{\dfrac{1+\gamma}{2}}\right]
\end{equation}
to which corresponds the value $R_0$ given by equation (3a). It 
is easy to verify that $r_0 > 2\eta$ and $R_0 > 0$. Now the 
singularity is beyond the throat and is smeared on a spherical 
surface, asymptotically large but not asymptotically flat, of 
radius $R(r) \to \infty$  as $r \to 2\eta$ and where also the 
red-shift and $G(r)$ become infinitely large [8,9].\par
When $\gamma = 1$ exactly, one has Schwarzschild solution of 
General Relativity
\begin{equation}
ds^2 = \dfrac{dr^2}{1-\dfrac{2M}{r}}+r^2 d\Omega^2- (1-
\dfrac{2M}{r}) dt^2
\end{equation}
where it appears a black-hole with event horizon at $R(r)=r 
=2M$. \par
If future measurements will establish that $\gamma \neq 1$ and 
if the Jordan frame is the physical frame, then the strong 
equivalence principle is violated (due to exotic matter in the 
case $\gamma > 1$) and it will possible to decide on the type of 
singularity  occurring. Moreover gravitation 
shall be better described by a suitable generalization of 
Einstein's theory. \par
Here we first assume $\gamma = 1$ exactly but then we introduce 
a fluctuation in the Schwarzschild metric by imposing a violation
of the strong equivalence principle, of the order of the Planck 
length $l_P$, which comes from the following variation of the 
gravitational radius : 
\begin{equation}
2\eta = 2M  \pm \dfrac{l_P}{2}
\end{equation}
Making use of Equation (5) one obtains for $\gamma$, at first 
order in $l_P/M$
\begin{equation}
\gamma = 1 \pm \frac{l_P}{M}
\end{equation}
so from Equation (4) the dimensionless coupling constant 
$\omega$  turns out to be $|\omega| \approx M/l_P$.
To make an example, in the case of a solar mass $M_{\bigodot}$ 
this would amount to a fluctuation of
$\Delta M_{\bigodot}/M_{\bigodot} \approx 10^{-38}$.\par
It may be useful to rewrite, in this approximation, the metric 
coefficients of the Brans-Dicke line element:
\begin{align*}
R^2(r) = r^2 \left[1-\dfrac {2 M \pm 
\dfrac{l_P}{2}}{r}\right]^{\mp \frac{3 l_p}{4 M}} \tag{12a}\\{}\\
e^{\mu(r)} =  \left[1-\dfrac {2 M \pm 
\dfrac{l_P}{2}}{r}\right]^{-1 \mp \frac{3l_P}{4 M}} 
\tag{12b}\\{}\\
e^{\nu(r)} =  \left[1-\dfrac {2 M \pm 
\dfrac{l_P}{2}}{r}\right]^{1\mp \frac{l_P}{4 M}} \tag{12c} 
\end{align*} 
\setcounter{equation}{12}
Here the upper sign refers to the case $\gamma >1$ and the lower 
one to the case $\gamma <1$. It is apparent that, while in the 
case $\gamma = 1$  the event horizon is fixed at $r_S = 2M$, in 
the presence of space-time fluctuations one may have, taking the 
extreme values of $\gamma$ given by Equation (11), either a naked
singularity at $r_< = 2M - l_P/2$ if $\gamma = 1 - l_P/M$ or a 
wormhole with a throat at $r_0 = 2M + (5/4)\, l_P$ if $\gamma = 1
+ l_P/M$; in this latter case we have beyond the throat at $r_> =
2M + l_P/2$ another singularity which is asymptotically large but
not asymptotically flat. We recall that $r_<$ and $r_>$ are the 
limits between which we required to fluctuate the metric. 
When $r$ exceeds $2M$ by several units of $l_P$ the 
solution of the field equations behaves as in the Schwarzschild 
case, but when $r$ varies around $2M$ in an interval 
containing all the possible singularities one is faced with 
different interchanging behaviours which are unpredictable on 
classical grounds. In this latter case the proper distance from 
the singularity of a point near to $2M$ is proportional to 
$\sqrt{M l_P}$. To make a numerical example in the case $\gamma 
>1$, the proper radial distance of $r_>$ (to which corresponds 
an infinite value of the standard radial coordinate $R$) from 
$r_0$ (to which corresponds the location $R_0$ of the throat) is
\begin{equation} 
L = \int_{r_>}^{r_0} e^{\mu(r)/2}dr \approx 
2.51 \sqrt{M l_P}
\end{equation}
and the proper volume comprised between the radii is
\begin{equation}
V = 4\pi \int_{r_>}^{r_0} R^2(r) e^{\mu(r)/2}dr \approx
36 \pi \sqrt{M^5 l_P} 
\end{equation} 
Numerical values for a solar mass are  $L = 3.9\, 10^{-16}\, m$ 
and $V = 3.4\, 10^{-10}\,m^3$. So, if it is  
correct to introduce  space-time fluctuations by means of the 
Brans-Dicke scalar field, then it would be needed a full quantum 
theory of gravitation to describe not only the kind and the 
evolution of the singularities but also the space-time that 
surrounds them at distances much greater than the  Planck 
length. If the scenario we have proposed here is realistic, 
it should turn out to be a good candidate both as a source of 
strong gravity waves and as a central engine required for the 
production of gamma ray bursts. 

\end{document}